\documentclass{iau}

\usepackage{amsmath}
\usepackage{graphicx}
\usepackage{multirow}

\begin{document}

\lefttitle{Silva-Andrade \& Souto}
\righttitle{A metallicity scale derived from Fe I and FeH lines in the APOGEE M dwarf spectra}


\jnlPage{1}{7}
\jnlDoiYr{2025}
\doival{10.1017/xxxxx}
\volno{395}
\pubYr{2025}
\journaltitle{Stellar populations in the Milky Way and beyond}

\aopheadtitle{Proceedings of the IAU Symposium}
\editors{J. Mel\'endez,  C. Chiappini, R. Schiavon \& M. Trevisan, eds.}

\title{A metallicity scale derived from Fe I and FeH lines in the APOGEE M dwarf spectra}

\author{Anderson Silva-Andrade$^1$ and Diogo Souto$^1$}
\affiliation{1 - Departamento de F\'isica, Universidade Federal de Sergipe, Av. Marcelo Deda Chagas, S/N, 49107-230, S\~ao Crist\'ov\~ao, SE, Brazil}

\begin{abstract}
We present metallicities derived from a sample of eleven M dwarfs belonging to wide binary systems with warmer FG primary companions observed by the high-resolution (R=22,500) near-infrared SDSS-IV APOGEE spectra. Using a plane-parallel one-dimensional local thermodynamic equilibrium (LTE) abundance analysis, we determine effective temperatures ($T_{\rm eff}$) based on the abundance equilibrium from the Fe I, FeH, OH, and H$_2$O spectral lines. We obtained three $T_{\rm eff}$ scales based on these lines and found that, regardless of the chosen $T_{\rm eff}$ scale, the M dwarf metallicities agree well with those of the warmer primaries, where the upper mean abundance difference limit is 0.04 $\pm$ 0.06. This good agreement confirms that FeH lines are a reliable indicator of $T_{\rm eff}$ in the $H-$band spectra.
\end{abstract}

\begin{keywords}
\href{http://astrothesaurus.org/uat/1558}{Spectroscopy}; 
\href{http://astrothesaurus.org/uat/154}{Binary systems};
\href{http://astrothesaurus.org/uat/555}{Fundamental parameters}
\end{keywords}

\maketitle

\section{Introduction}
M dwarfs represent a substantial portion of the stellar population in the Milky Way (\citealt{Salpeter1955}; \citealp{Miller}; \citealp{Henry}). They are characterized by relatively low temperatures spanning from 2500K to 3900K and possess atmospheres abundant in molecules like H$_2$O, TiO, and CO, which prominently feature in their optical spectra (\citealp{Souto2020}). Utilizing high-resolution spectroscopic techniques in the near-infrared (NIR) provides means to ascertain key parameters of these stars, such as their effective temperature ($T_{\rm eff}$), surface gravity (log g), and individual elemental abundances, as molecular blending is less pronounced in this spectral range (\citealp{Allard2000}). In this work, our goal is to derive a scale of effective temperature ($T_{\rm eff}$), metallicity ([Fe/H]), and surface gravity (log g) based on iron and oxygen abundances, using neutral Fe I and molecular FeH, OH, and H$_2$O transitions in the APOGEE spectra for a sample of cool dwarf stars.

\section{Sample and methodology}

We adopted the near-infrared $H-$band SDSS-IV APOGEE DR17 (\citealp{Abdurro'uf}) spectra. APOGEE is a cryogenic high-resolution (R$\sim$22,500) multi-fiber spectrograph operating in both hemispheres.
From the APOGEE wide binaries catalog from \cite{El}, we selected eleven wide binaries M dwarf stars, each with a respective primary FG-type companion (5000 K $<$ $T_{\rm eff}$ $<$ 6000 K). 
For deriving the atmospheric parameters for M dwarfs, we used the Turbospectrum code (\citealp{Plez}), the MARCS atmospheric models (\citealp{Gustafsson}), and the APOGEE DR17 line list (\citealp{Smith}) to generate spectral synthesis and derive $T_{\rm eff}$, log g's, and metallicities using Fe I, FeH, OH, and H$_2$O lines. This procedure is described in more detail in \cite{Souto2020}. 
We adopted the calibrated atmospheric parameters from ASPCAP DR17 for the warmer primaries and determined their abundances by comparing the observed spectra with synthetic spectra. The final abundances correspond to the fits that minimize the chi-square value while also providing a visually reasonable adjustment.


\section{Results and discussions}

We obtain three $T_{\rm eff}$ scales for the M dwarfs studied in this work, one using the Fe I and FeH lines (pair $T_{\rm eff}$--A(Fe)), the other with the OH and H$_2$O lines (pair $T_{\rm eff}$--A(O)) and the third one, which is a mean value for the former ones. 
We find the pair $T_{\rm eff}$--A(Fe) having $T_{\rm eff}$ values systematically higher than those from the $T_{\rm eff}$--A(O) pair by about 96K. This offset was initially observed by \cite{Souto2020}, where the authors suggested that it could indicate problems in the log gf value adopted from the FeH linelist from \cite{Hargreaves}. 

\begin{figure}[ht!]
    \centering
     \includegraphics[width=0.49\linewidth]{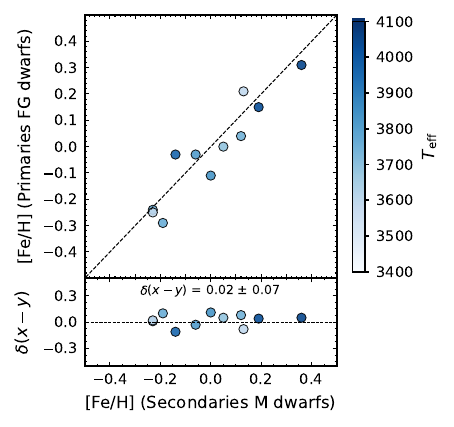}
     \includegraphics[width=0.49\linewidth]{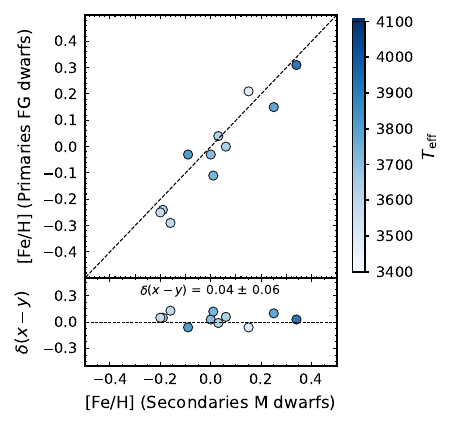}
     \includegraphics[width=0.49\linewidth]{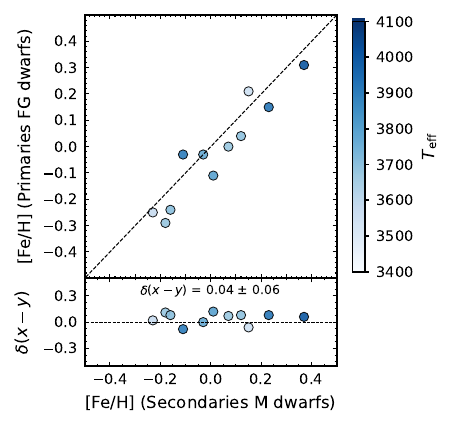}
    \caption{The metallicity--metallicity diagrams for secondary M and warmer primaries FG dwarfs. The panels in the upper-left and upper-right corners illustrate the determinations using the $T_{\rm eff}$--A(Fe) and $T_{\rm eff}$--A(O) pairs, respectively. The bottom panel shows the case using the mean value from both pairs. Each panel includes a residual diagram at the bottom.}
  \label{fig:effectivetemp}
\end{figure}

In this work, we can compare the metallicity scales from the M dwarfs using the iron abundance indicator from their spectral lines of FeH or Fe I. For the FG-dwarfs, we rely only on Fe I lines.
We find that, regardless of the adopted $T_{\rm eff}$ scale for the M dwarfs, we have an excellent agreement between the metallicities derived from the M and FG dwarfs. 
Figure \ref{fig:effectivetemp} illustrates the one-to-one metallicity comparison for M dwarfs (on the x-axis) and FG dwarfs (on the y-axis). 
The first panel, in the upper-left corner, shows the metallicity determinations for M dwarfs based on the effective temperature derived from the $T_{\rm eff}$--A(Fe) pair, while the second panel, in the upper-right corner, presents the determinations using the $T_{\rm eff}$--A(O) pair. The bottom panel illustrates the case where the mean value obtained from both pairs was used.
As seen in the figure, the results from the three different $T_{\rm eff}$ and metallicity scales are indistinguishable, especially when considering the typical scatter and uncertainty (the typical metallicity uncertainty is about 0.10 dex). Therefore, we can conclude that the log gf values taken from the APOGEE line list for FeH are as reliable as those obtained from the Fe I lines. The results confirm the reliability of our $T_{\rm eff}$ and metallicity scale. This good agreement validates the methods employed and reinforces the accuracy of the atmospheric parameters derived for M dwarfs.

\section*{Acknowledgements}
A.S-A. thank the support from fellowship by Coordenação de Aperfeiçoamento de Pessoal de Ensino Superior – CAPES. D.S. thank the National Council for Scientific and Technological Development – CNPq process No. 404056/2021-0.  

Funding for the Sloan Digital Sky Survey IV has been provided by the Alfred P. Sloan Foundation, the U.S. Department of Energy Office of Science, and the Participating Institutions. SDSS-IV acknowledges support and resources from the Center for High-Performance Computing at the University of Utah. The SDSS website is www.sdss.org.
SDSS-IV is managed by the Astrophysical Research consortium for the Participating Institutions of the SDSS Collaboration including the Brazilian Participation Group, the Carnegie Institution for Science, Carnegie Mellon University, the Chilean Participation Group, the French Participation Group, Harvard-Smithsonian Center for Astrophysics, Instituto de Astrof\'isica de Canarias, The Johns Hopkins University, 
Kavli Institute for the Physics and Mathematics of the Universe (IPMU) /  University of Tokyo, Lawrence Berkeley National Laboratory, Leibniz Institut f\"ur Astrophysik Potsdam (AIP),  Max-Planck-Institut f\"ur Astronomie (MPIA Heidelberg), Max-Planck-Institut f\"ur Astrophysik (MPA Garching), Max-Planck-Institut f\"ur Extraterrestrische Physik (MPE), National Astronomical Observatory of China, New Mexico State University, New York University, University of Notre Dame, Observat\'orio Nacional / MCTI, The Ohio State University, Pennsylvania State University, Shanghai Astronomical Observatory, United Kingdom Participation Group,
Universidad Nacional Aut\'onoma de M\'exico, University of Arizona, University of Colorado Boulder, University of Oxford, University of Portsmouth, University of Utah, University of Virginia, University of Washington, University of Wisconsin, Vanderbilt University, and Yale University.

\end{document}